\documentclass[aps,prd,amsmath,twocolumn,showpacs]{revtex4}

\usepackage{epsfig}

\usepackage{epstopdf}

\usepackage{graphics}
\usepackage{latexsym}
\usepackage{amsmath}
\usepackage{amssymb}
\usepackage{array,booktabs,makecell}
\usepackage{rotating}
\usepackage{subfigure}
\usepackage{bm}
\usepackage{xcolor}

\usepackage{hyperref}
\usepackage{cleveref}
\begin{document}

\title{Nucleon axial form factor from generalized parton distributions}

\author{Hadi Hashamipour$^{a}$}
\email{h\_hashamipour@sbu.ac.ir}

\author{Muhammad Goharipour$^{b}$}
\email{muhammad.goharipour@ipm.ir}

\author{Siamak S. Gousheh$^{a}$}
\email{ss-gousheh@sbu.ac.ir}
\affiliation {
$^{a}$Department of Physics, Shahid Beheshti University, G.C., Evin, Tehran 19839, Iran  \\
$^{b}$School of Particles and Accelerators, Institute for Research in Fundamental Sciences (IPM), PO Box 19568-36681, Tehran, Iran}

\date{\today}

\begin{abstract}

It is well established that the nucleon form factors can be related to Generalized Parton Distributions (GPDs) through sum-rules. On the other hand, GPDs can be expressed in terms of Parton Distribution Functions (PDFs) according to Diehl's model. In this work, we use this model to calculate polarized GPDs for quarks ($\widetilde{H}_q$) using the available polarized PDFs obtained from the experimental data, and then study the axial form factor of nucleon. We determine parameters of the model using standard $\chi^2$ analysis of experimental data. It is shown that some parameters should be readjusted, as compared to some previously reported values, to obtain better consistency between the theoretical predictions and experimental data. Moreover, we study in details the uncertainty of nucleon axial form factor due to various sources.

\end{abstract}


%
\maketitle

\section{Introduction}\label{sec:one} 
Measurement of proton spin asymmetry in polarized deep inelastic scattering (DIS) experiment performed by European Muon Collaboration in 1987 \cite{Ashman:1987hv} showed that the total contribution from quark spin to the proton's spin is less than half. Since that time, along with the studies of unpolarized parton distribution functions (PDFs), many experimental, theoretical and phenomenological investigations have been performed to understand the constituents of the proton's spin and also determine their distributions. In this regard, the determination of polarized PDFs (PPDFs), which describe the structure of a nucleon in a helicity eigenstate, by performing a global analysis of available experimental data has also been of great interest~\cite{deFlorian:2009vb,Blumlein:2010rn,Leader:2010rb,deFlorian:2014yva,Nocera:2014gqa,
Jimenez-Delgado:2014xza,Sato:2016tuz,Shahri:2016uzl,Khanpour:2017cha,Ethier:2017zbq,Khanpour:2017fey,Salajegheh:2018hfs}.
On the other hand, the structure of nucleon both in unpolarized and polarized cases can be investigated in more detail using generalized parton distributions (GPDs)~\cite{Goeke:2001tz,Diehl:2003ny,Polyakov:2002yz,Freund:2002qf,Scopetta:2003et,Belitsky:2005qn,Boffi:2007yc,Guzey:2005ec,Guzey:2006xi,Hagler:2007xi,Alexandrou:2013joa,Kumericki:2007sa,Guidal:2013rya,Kumericki:2016ehc,Khanpour:2017slc} which comprise important concepts of Quantum Chromodynamics (QCD). Actually, GPDs provide quantitative information on the longitudinal and transverse distribution of partons inside the nucleon, and also their intrinsic and orbital angular momenta. Therefore, studying GPDs can shed light on various aspects of hadron structure.

An accurate knowledge of GPDs is essential for understanding and describing various hard exclusive processes~\cite{Goeke:2001tz,Diehl:2003ny,Belitsky:2005qn}, such as deeply virtual Compton scattering (DVCS), timelike Compton scattering (TCS), exclusive meson production by longitudinally polarized photons and photoproduction of heavy vector mesons~\cite{Ivanov:2004vd}. One of the main important properties of GPDs is their mutual relations with PDFs and elastic form factors (FFs). On one hand, GPDs are the off-forward kinematic generalizations of the ordinary PDFs which play a crucial role in inclusive DIS. In other words, PDFs can be recovered from GPDs (in the so-called forward limit) by setting to zero the extra variables in GPDs, such as the transverse momentum between the initial and final protons and skewness parameter. On the other hand, FFs (including the electric and magnetic form factors, or even the form factors associated with the energy-momentum tensor) which are other important quantities giving us valuable information on the structure of nucleon can be obtained from GPDs as well~\cite{Diehl:2004cx,Diehl:2007uc,Diehl:2013xca}. In fact, GPDs give us not only all the information contained in FFs, but also other useful information, for example, about the transverse displacement of partons~\cite{Burkardt:2002ks}. Consequently, from this point of view, GPDs are generalization of both PDFs and FFs. It is worth noting that the axial form factors (AFFs) which are fundamental quantities that describe spin content of the nucleon are also intimately related to polarized GPDs (see~\Cref{sec:two}).

Just like PDFs and PPDFs, GPDs are essentially non-perturbative objects, so that they cannot be determined directly from the perturbative QCD apart from first Mellin moments in special cases in lattice QCD~\cite{Hagler:2007xi,Alexandrou:2013joa}. Although early studies of GPDs using various dynamical models of the nucleon structure (see Ref.~\cite{Khanpour:2017slc} and references therein) have played an important role for better understanding of GPDs and exclusive processes, but at the moment, more attention is being paid to determine GPDs from fitting the available experimental data (see Ref.~\cite{Kumericki:2016ehc} and references therein). Actually, the extraction of GPDs from exclusive processes, for which all particles are detected in the final state, is theoretically well developed.  There are valuable experimental data from DVCS on proton at HERA collider which cover a wide kinematical region (see Table 4 of~\cite{Kumericki:2016ehc}). For the case of DVCS experiments with fixed proton target, there are also some data on various observables from HERMES, CLAS and Hall A Collaborations (see Table 5 of~\cite{Kumericki:2016ehc}). Nevertheless, DVCS data-taking will be continued at CERN by the COMPASS Collaboration~\cite{dHose:2004usi}. The future measurements in Jefferson Lab (JLab) by CLAS Collaboration~\cite{Armstrong:2017wfw} with experiments starting at 12 GeV (CLAS12) will also provide new information on valence region. Moreover, one of the main goals of future Electron-Ion Collider (EIC) is the measurement of DVCS observables and FFs~\cite{Accardi:2012qut} that makes the extraction of both $H$ and $E$ GPDs possible.

As mentioned, FFs can be written in terms of GPDs, and hence their measurements can give us useful important information on GPDs. In the case of nucleon spin studies, AFFs which are related to polarized GPDs can be extracted using various approaches~\cite{Tsushima:1988xv,JuliaDiaz:2004qr,Mamedov:2016ype,Liu:2016kpb,Anikin:2016teg,Adamuscin:2007fk,Aznauryan:2012ba,Ramalho:2017tga}.
One can find a review of experimental data in Refs.~\cite{Bernard:2001rs,Schindler:2006jq}. Many lattice QCD calculations of FFs have also been done since 1980s and have lead to considerable results. In recent years, lattice QCD simulations of AFFs have been presented for pion masses in the range $m_\pi=0.2-0.6$ GeV~\cite{Bhattacharya:2013ehc,Liang:2016fgy,Green:2017keo,Yao:2017fym,Abdel-Rehim:2015owa,Bali:2014nma,Bhattacharya:2016zcn}. Very recently, PACS Collaboration has reported the result of a lattice QCD calculation of nucleon AFF in 2+1 flavor near the physical pion mass~\cite{Ishikawa:2018rew}. In addition, neural networks can be applied to extract nucleon AFF from  experimental data. In particular, the authors of Ref.~\cite{Alvarez-Ruso:2018rdx} have used this tool to analyze the neutrino-deuteron scattering data measured by the Argonne National Laboratory (ANL) bubble chamber experiment.

In this work, we study the nucleon axial form factor and polarized GPDs, given the fact that they are connected via sum rules. Although, there are various models~\cite{Pasquini:2005dk,Pasquini:2006dv,Dahiya:2007mt,Frederico:2009fk,Mukherjee:2013yf,Maji:2015vsa} and parameterizations~\cite{Goldstein:2010gu,Goldstein:2013gra,Sharma:2016cnf} for GPDs, we use a practical ansatz suggested by Diehl~\cite{Diehl:2004cx,Diehl:2007uc,Diehl:2013xca} which relates the predetermined (polarized) PDFs as input to (polarized) GPDs. An important advantage of this ansatz is that it has a few free parameters to be fixed by analyzing experimental data. Considering different scenarios, we determine parameters of the model using standard $\chi^2$ analysis of experimental data for nucleon AFF.

This paper is organized as follows. In \Cref{sec:two}, the theoretical framework of our study is presented and we briefly describe the physics related to GPDs and AFFs. Our method to obtain optimum values for the polarized GPDs of quarks using the available experimental data for nucleon AFF is also introduced in this section. \Cref{sec:three} is devoted to introduction of the experimental data which are used in our $ \chi^2 $ analyses. In \Cref{sec:four}, we study in detail the nucleon AFF with emphasis on its dependence on PPDFs according to Diehl model~\cite{Diehl:2004cx,Diehl:2007uc,Diehl:2013xca}, and also the value of scale $ \mu^2 $ associated with the PPDFs. Moreover, we investigate the model uncertainties that are imposed on the nucleon AFF from various sources. In \Cref{sec:five}, we determine the best values of parameters of the model by performing some $ \chi^2 $ analyses of nucleon AFF data in various scenarios and discuss the results obtained and possible outlooks.
Finally, we summarize our results and conclusions in \Cref{sec:six}.\\

%
\section{Theoretical framework}\label{sec:two}

In this section, we briefly review physical concepts on GPDs and nucleon AFF, and present the theoretical framework we use to obtain optimum values and bounds for polarized GPDs using the available experimental data for nucleon AFF. As mentioned in the Introduction, GPDs (PDFs) are non-perturbative objects needed for describing hard exclusive (inclusive) electroproduction processes, which are defined as matrix elements of quark and gluon operators at a light-like separation between two proton states with different (same) momenta. GPDs are also universal objects just like PDFs, because they can be defined in the framework of QCD collinear factorization for hard exclusive processes~\cite{Collins:1998be,Collins:1996fb} such as DVCS and exclusive meson production by longitudinally polarized photons. The importance of GPDs is due to the fact that they contain valuable information on the hadron structure in QCD. Actually, the distributions of quarks and gluons in hadrons in terms of both momentum fractions and position in the transverse plane can be well described through GPDs.

In the present work, we use the convention of Ji~\cite{Ji:1996ek} for GPDs, in which $H$, $E$, $\widetilde{H}$ and $\widetilde{E}$ are defined as~\cite{Belitsky:2005qn,Diehl:2003ny}:
\begin{widetext}
\begin{flalign}
\label{Eq:1}
& \frac{1}{2}\int \frac{d z^-}{2\pi}\, e^{ix P^+ z^-}
\langle p'|\, \bar{q}(-\frac{1}{2} z)\, \gamma^+ q(\frac{1}{2} z) 
\,|p \rangle \Big|_{\substack{z^+=0,\\\bm{z_\perp=0}}}
= \frac{1}{2P^+} \left[
H^q(x,\xi,t)\, \bar{u}(p') \gamma^+ u(p) +
E^q(x,\xi,t)\, \bar{u}(p') 
\frac{i \sigma^{+\alpha} \Delta_\alpha}{2m} u(p)
\, \right] ,
\nonumber \\
&\frac{1}{2} \int \frac{d z^-}{2\pi}\, e^{ix P^+ z^-}
\langle p'|\, 
\bar{q}(-\frac{1}{2} z)\, \gamma^+ \gamma_5\, q(\frac{1}{2} z)
\,|p \rangle \Big|_{\substack{z^+=0,\\\bm{z_\perp=0}}}
= \frac{1}{2P^+} \left[
\widetilde{H}^q(x,\xi,t)\, \bar{u}(p') \gamma^+ \gamma_5 u(p) +
\widetilde{E}^q(x,\xi,t)\, \bar{u}(p') \frac{\gamma_5 \Delta^+}{2m} u(p)
\, \right],
\end{flalign}
\end{widetext}
where $z=\left(z^+,\bm{z_\perp},z^-\right)$. As one can readily see from Eq.~(\ref{Eq:1}), GPDs have three degrees of freedom, and then are expressed as functions of three parameter $ x $, $ \xi $ and $ t $. The first argument is the well-known Bjorken scaling variable (the average momentum fraction) $x=\frac{Q^2}{2 p\cdot q}$, with photon virtuality $ Q^2 $. Another longitudinal variable that has a crucial role in GPDs is $\xi=\frac{p^+-p'^+}{p^++p'^+}$, which is called ``skewness". The last argument is $t=(p'-p)^2=\Delta^2= -Q^2$, i.e. the squared of the momentum transferred to the target.

As mentioned, GPDs cannot be calculated from perturbative QCD, but there are some lattice QCD calculations~\cite{Hagler:2007xi,Alexandrou:2013joa}. A suitable method to extract GPDs  
is performing a $ \chi^2 $ analysis of experimental data using factorization theorem. Hard exclusive processes such as DVCS~\cite{Kumericki:2016ehc} and meson production~\cite{Diehl:2013xca} are most used experiments for the extraction of GPDs. As data on hard exclusive processes are much less than inclusive processes, extraction of GPDs from experimental data is not yet feasible with precisions comparable to PDFs. One of the best ways to overcome this problem is using a model for GPDs, but with as few parameters as possible. In this work, we implement Diehl's model \cite{Diehl:2004cx} for calculating polarized GPDs which can be expressed in terms of PPDFs and also has few free parameters to be fixed by fitting to nucleon AFF data. We describe the model below. 

It is well established now that various nucleon FFs can be related to GPDs through sum-rules~\cite{Diehl:2013xca}. For example, the Dirac and Pauli form factors, $ F_1 $ and $ F_2 $, for proton and neutron can be expressed in the following form 
\begin{align}
F_i^p= e_u F_i^u + e_d F_i^d + e_s F_i^s, \nonumber \\ 
F_i^n= e_u F_i^d + e_d F_i^u + e_s F_i^s, 
\label{Eq:2}
\end{align}
where $ i=1, 2 $ and $ F_i^q $ is the contribution from quark flavor $ q $ to the nucleon form factor $ F_i^A $, with $ A=p,n $. As usual, $ e_q $ is the electric charge of the quark in units of the positron charge. Now, the flavor form factors $ F_i^q $ can be written in terms of the proton ``valence GPDs" $ H_v $ and $ E_v $ for unpolarized quarks of flavor $ q $ as
\begin{align}
F_1^q(t)= \int_0^1 dx~H_v^q(x,t), \nonumber \\ 
F_2^q(t)= \int_0^1 dx~E_v^q(x,t), 
\label{Eq:3}
\end{align}
where valence GPDs $ {\cal G}_v= H_v, E_v $ for flavor $ q $ are expressed in terms of ``quark GPDs" $ {\cal G} $ as
\begin{equation}
{\cal G}_v^q(x,t)= {\cal G}^q (x,\xi=0,t) + {\cal G}^q (-x,\xi=0,t),
\label{Eq:4}
\end{equation}
with $ {\cal G}^q (-x,\xi=0,t)= - {\cal G}^{\bar q} (x,\xi=0,t) $. Note that the result, as a consequence of Lorentz invariance, is independent of skewness $\xi$, so one can choose zero skewness GPDs and omit this variable.

As we pointed out, some models for GPDs use ordinary PDFs as input. Considering this fact, PDFs can be defined as 
\begin{flalign}
{\label{Eq:5}}
q(x)=\int \frac{dz^-}{2\pi}e^{-ixP^+z^-} \langle p|\, \bar{q}(z)\, \gamma^+ q(0) 
\,|p \rangle \Big|_{\substack{z^+=0,\\\bm{z_\perp}=0}},
\end{flalign}  
and then recovered from GPDs at forward limit ($ t=0 $). For example, for positive $ x $, the GPD $ H $ changes to the usual quark and antiquark densities  as $ H^q(x,0,0)=q(x) $ and $ H^q(-x,0,0)=\bar{q}(x) $. According to the Diehl's ansatz~\cite{Diehl:2004cx} which gives $x$ and $t$ dependence of GPDs at zero skewness, the valence GPDs $ H_v^q $, for example, can be related to ordinary valence PDFs as
\begin{equation}
H_v^q(x,t)= q_v(x)\exp [tf_q(x)],
\label{Eq:6}
\end{equation}
in which the profile functions $ f_q(x) $ specifies the $ x $ dependent width. Actually, this ansatz assumes an exponential $ t $ dependence with a $ x $-dependent slope for $ H_v^q $. The profile functions $ f_q(x) $ can have the simple form shown below, which we shall henceforth call the simple ansatz,
\begin{equation}
f_q(x)=\alpha^{\prime}(1-x)\log \frac{1}{x}.
\label{Eq:7}
\end{equation}
This ansatz, along with a more complex one also given in \cite{Diehl:2004cx}, were used for example, in Ref.~\cite{Diehl:2007uc} for the strange Dirac form factor $ F_1^{s} $. The value of $ \alpha^{\prime} $ can be extracted by analyzing the soft hadronic scattering processes like kaon-nucleon scattering or photoproduction of the mesons; Various analyses have indicated that its values should be close to one~\cite{Diehl:2004cx,Diehl:2007uc,Diehl:2013xca}.

In analogy with the Dirac and Pauli FFs, the nucleon axial form factor can be expressed in terms of polarized GPDs as~\cite{Diehl:2013xca}  
\begin{flalign}
G_A(t)=&\int_0^1 dx \left[\widetilde{H}^u_v(x,t)-\widetilde{H}^d_v(x,t)\right]+\nonumber\\
2&\int_0^1 dx \left[\widetilde{H}^{\bar{u}}(x,t)-\widetilde{H}^{\bar{d}}(x,t)\right].
\label{Eq:8}
\end{flalign}
Note that, for valence polarized GPDs $\widetilde{H}^q_v$, we have
\begin{equation}
\label{Eq:9}
\widetilde{H}^q_v(x,t)\equiv \widetilde{H}^q(x,\xi=0,t)-\widetilde{H}^q(-x,\xi=0,t),
\end{equation}
with 
$ \widetilde{H}^q(-x,\xi=0,t)= \widetilde{H}^{\bar{q}}(x,\xi=0,t) $. In fact, one can write the quark contribution to AFF generally as an integral of polarized GPDs over Bjorken $x$,
\begin{equation}
 G_A^q(t)=\int_{0}^{1} dx~\widetilde{H}^q(x,t),
\label{Eq:10}
\end{equation}
where $ q $	covers here both valence and sea type contributions of $ up $ and $ down $ quarks. To be more precise, Eq.~(\ref{Eq:8}) clearly shows that, in contrast to Pauli and Driac FFs, the axial form factor contains also some contributions from the sea quark sector. Although these contributions are not significant compared with those come from valence sector, they cannot be neglected. It is worth noting that Eq.~(\ref{Eq:10}) is also the intrinsic spin contribution of quark $q$ to the spin of nucleon. 

According to Diehl's model, an ansatz similar to that shown in Eq.~(\ref{Eq:6}) can be also considered for valence polarized GPDs $\widetilde{H}^q_v$, so that they can be related to valence polarized PDFs, $\Delta q_v(x)\equiv q^+(x)-q^-(x)$, as following
\begin{equation}
\widetilde{H}^q_v(x,t)=\Delta q_v(x) \exp [t \widetilde{f}_q(x)],
\label{Eq:11}
\end{equation}
where $\widetilde{f}_q(x)$ is the corresponding profile function which can have again a simple form like Eq.~(\ref{Eq:7}), or a complex form with more adjustable parameters. For simplicity we use the ansatz Eq.~(\ref{Eq:11}) both for $ \widetilde{H}_v^q(x,t) $ and $ \widetilde{H}^{\bar q}(x,t) $ in Eq.~(\ref{Eq:8}). In fact, this is an ad hoc ansatz for $ \widetilde{H}^{\bar q}(x,t) $ whose physical motivation is not as strong as that of the dominant $ \widetilde{H}_v^q(x,t) $.

%
\section{Experimental data}\label{sec:three}

One of the best ways to investigate the electromagnetic and weak structure of hadrons is using the electroweak probes and then measuring various structure form factors. Actually, the extraction of the electromagnetic nucleon FFs has a long history and remains a popular field of experimental research. An overview and discussion of FF data can be found in~\cite{Diehl:2013xca}. Although the vector electroweak FFs, which give us valuable information on the spatial distribution of the charge and magnetism, have been explored experimentally to a large extent, our information about the axial form factors is very limited. At the present, there are only two classes of experiments that can be used to determine AFF: first, (anti)neutrino scattering off protons or nuclei, and second, charged pion electroproduction. 

In this section, we introduce the nucleon AFF data that is used in our study. For a clear and thorough review and discussion of AFF data, one can refer to Refs.~\cite{Bernard:2001rs,Schindler:2006jq}).  
Reference~\cite{Bernard:2001rs} also includes clear explanations about the relevant methods to determine AFF of the nucleon. For the case of (anti)neutrino scattering experiments, we use the data obtained by analyzing the measurements of (quasi)elastic (anti)neutrino scattering off Ca, O and Fe nuclei from MiniBooNE experiments~\cite{Butkevich:2013vva}. These data cover a wide range of $ Q^2 $ in the interval $ 0.25 < Q^2 < 0.9 $ GeV$ ^2 $. As mentioned, the other information on the AFF is obtained from the analysis of charged pion electroproduction off protons, slightly above the pion production threshold. Although such type of analysis is more complicated, but there are more experimental data of this class. In the present work, we use a wide range of charged pion electroproduction data~\cite{Bernard:2001rs,Amaldi:1970tg,Amaldi:1972vf,Bloom:1973fn,Brauel:1973cw,DelGuerra:1975uiy,DelGuerra:1976uj,Joos:1976ng,Esaulov:1978ed,Choi:1993vt,Choi:1993}.
In such analyses, the Nambu, Lurié and Shrauner low-energy (NLS) theorem~\cite{Nambu:1997wa,Nambu:1997wb} is firstly used for the electric dipole amplitude $E^{(-)}_ {0+}$ at production threshold. Note that the NLS theorem is valid for soft pions, namely pions that have vanishing four-momentum. Then, the so-called hard pion corrections (model-dependent corrections), labeled as SP, FPV, DR and BNR, are used to connect the low-energy theorem to the data; in other words, to the realistic case with a finite pion mass~\cite{Bernard:2001rs}.

In most cases, the AFF data are presented as a simple parametrization~\cite{Bernard:2001rs}. The commonly used parametrization for AFF is the so-called dipole ansatz, for its $Q^2$ dependence, which is as follows:
\begin{equation}
\label{Eq:12}
G^{\textnormal{dipole}}_A(Q^2)=\frac{g_A}{1+\frac{Q^2}{M_A^2}},
\end{equation}
where the value of axial mass $M_A$ varies between $1.03$ and $1.07 $ GeV depending on the method which is used for analyzing experimental data~\cite{Bernard:2001rs,Schindler:2006jq}. The value of $G_A$ at $t=0$, the axial charge $ g_A $, is precisely determined from  $\beta$-decay experiments. As can be seen, Eq.~(\ref{Eq:12}) has only a single free parameter, $M_A$, which should be fixed by fitting the experimental data. It should be also noted that, in the Breit frame and for small momenta, such $Q^2$ dependence of AFF leads to an exponentially decrease for the axial charge distribution~\cite{Alvarez-Ruso:2018rdx}.
However, from a theoretical point of view, it has been indicated that this ansatz is not a good choice, e.g. see~\cite{Bhattacharya:2011ah,Bhattacharya:2015mpa}. For example, a recent analysis of $G_A$~\cite{Meyer:2016oeg}, using conformal mapping or z-expansion, shows that the dipole ansatz systematically underestimates the uncertainty of AFF. Therefore, in this work, we do not implement dipole ansatz and use the experimental data points directly.

Another important point which should be noted is that the experimental data of Refs.~\cite{Butkevich:2013vva,Amaldi:1970tg,Amaldi:1972vf,Bloom:1973fn,Brauel:1973cw,DelGuerra:1975uiy,DelGuerra:1976uj,Joos:1976ng,Esaulov:1978ed,Choi:1993vt,Choi:1993} for AFF have been presented as ratio to $G_A$ at $t=0$, i.e. $ G_A(Q^2)/G_A(0) $. Hence one can use two approaches for analyzing these data: 1) using the original data as ratios, and 2) using data as $ G_A(Q^2) $. In the next section, we first use both of them and compare their results, and then continue our investigations with just $ G_A(Q^2) $ data.
Note that for extracting  $ G_A(Q^2) $ data from original $ G_A(Q^2)/G_A(0) $ data we need the value of $ G_A(0) $ (axial charge $ g_A $). Although more accurate results for $ g_A $ can be extracted from recent measurements of the nucleon lifetime~\cite{Gonzalez-Alonso:2018omy},  we use the latest value from PDG~\cite{Tanabashi:2018oca}, i.e. $g_A=1.2723\pm 0.0023$. 

As a last point, note that the total number of data points from Refs.~\cite{Butkevich:2013vva,Amaldi:1970tg,Amaldi:1972vf,Bloom:1973fn,Brauel:1973cw,DelGuerra:1975uiy,DelGuerra:1976uj,Joos:1976ng,Esaulov:1978ed,Choi:1993vt,Choi:1993} that we can use in our study is 84. However, comparing data points from various experiments, one find that some of points have same $ Q^2 $. Therefore, another way to analyze these data is removing those with same value of $ Q^2 $ and retaining most accurate ones. If we do this, 40 data points will remain which we refer to them as ``reduced data", in the following.

%
%
\section{Study of nucleon axial form factor}\label{sec:four}

In the previous sections, we presented the theoretical framework and experimental information related to the nucleon axial form factor $ G_A $. In this section we study $ G_A $ in detail with emphasis on its dependence on PPDFs according to ansatz Eq.~(\ref{Eq:11}), and also the value of scale $ \mu$ at which PPDFs are chosen. Moreover, we investigate the model uncertainties that are imposed upon the nucleon AFF due to the PPDFs uncertainties and also variation of $ \alpha^{\prime} $ in profile functions $ f(x) $.

\subsection{Dependence of $G_A$ on the PPDFs}

As can be seen from Eq.~(\ref{Eq:8}), the nucleon axial form factor can be related to PPDFs through its dependence on polarized GPDs, and the relationship between polarized GPDs and PPDFs. It is natural to expect that using different sets of PPDFs to perform calculations, should not change behaviour and magnitude of the resulting $ G_A $, otherwise the model can be considered as not flexible enough or inconsistent. Consequently, in this section, we choose the simple ansatz given by Eq.~(\ref{Eq:7}) and calculate $ G_A $ using different sets of PPDF and compare the results to see if such dependence is present. We will show below that such dependence is not present and hence the model can be used to describe the data.

\begin{figure}[t!]
\centering
\includegraphics[width=0.53\textwidth]{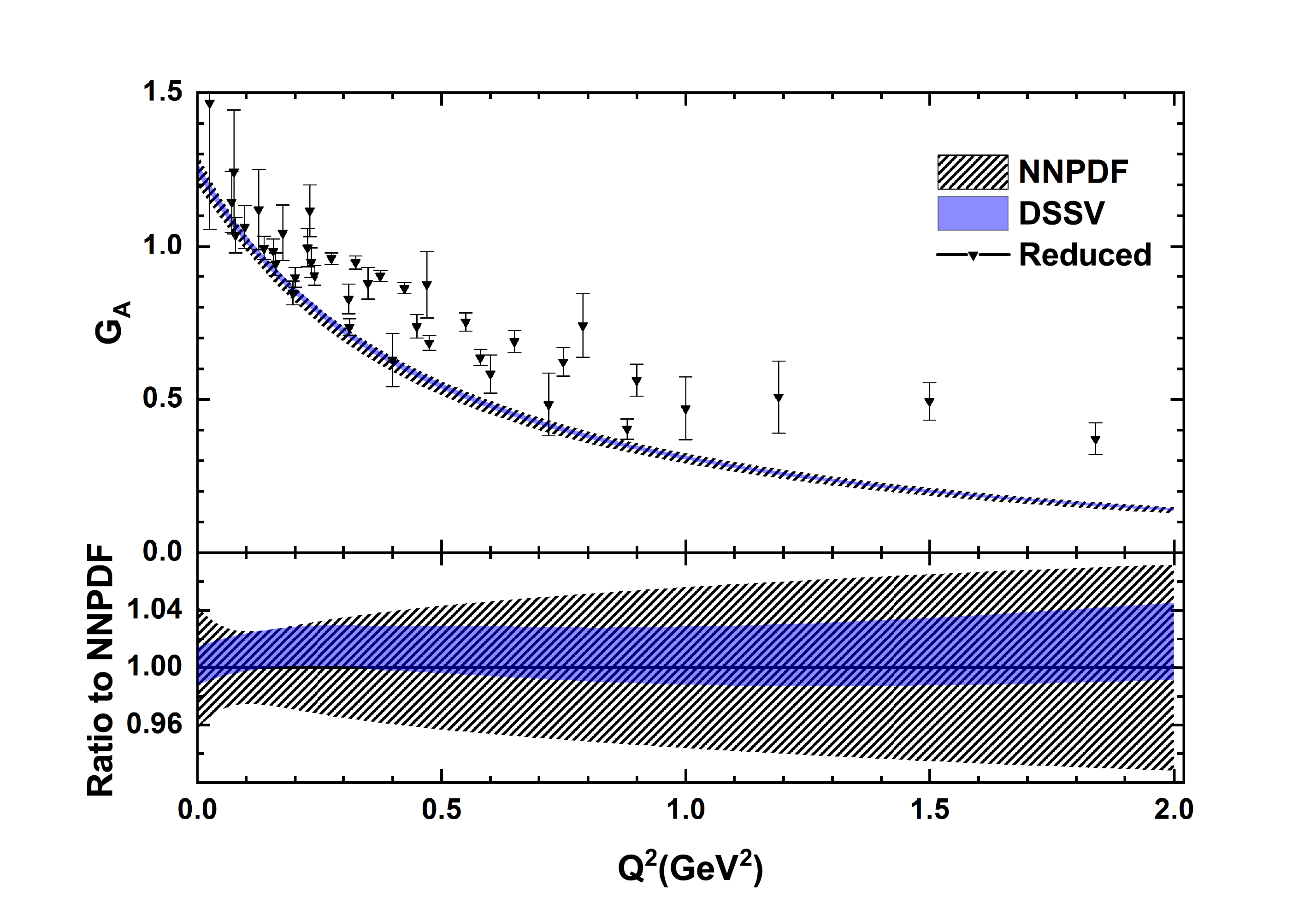}
\caption{The theoretical results obtained for nucleon AFF, $ G_A $, as a function of $ Q^2 $ using simple ansatz Eq.~(\ref{Eq:7}) with NLO PPDF sets \texttt{DSSV08}~\cite{deFlorian:2009vb} (filled band) and \texttt{NNPDFpol1.1}~\cite{Nocera:2014gqa} (hachured band) taken at $ \mu=2 $ GeV, value of $\alpha'$ is set to $0.95~\mathrm{GeV}^2$. The data points labeled as ``reduced" are related to various experiments selected in a procedure explained in Sec.~\ref{sec:three}.}
\label{fig:fig1}
\end{figure}

Since the evaluation of the uncertainty in $ G_A $ due to the PPDF uncertainties is also of interest, we choose two NLO PPDF sets \texttt{DSSV08}~\cite{deFlorian:2009vb} and \texttt{NNPDFpol1.1}~\cite{Nocera:2014gqa} which provide error PPDF sets in addition to their central fit results \footnote{The \texttt{NNPDFpol1.1} PPDFs are available through the \texttt{LHAPDF} package~\cite{Buckley:2014ana}.}. An important advantage of these sets is that one can calculate the uncertainties in any quantity related to PPDFs more easily~\cite{Pumplin:2001ct,Nadolsky:2008zw}. 
Figure~(\ref{fig:fig1}) shows the results obtained for nucleon axial form factor $ G_A $ as a function of $ Q^2 $ in which the hachured and filled bands correspond to the \texttt{NNPDFpol1.1} and \texttt{DSSV08} PPDFs, respectively. In order to investigate in more detail the differences between the predictions in various regions of $ Q^2 $, their ratios to the \texttt{NNPDFpol1.1} prediction have also been plotted in the bottom panel of Fig.~(\ref{fig:fig1}). Moreover, the experimental data from various experiments, which as explained in the previous section are referred to as ``reduced", have been shown for comparison. Both \texttt{DSSV08} and \texttt{NNPDFpol1.1} PPDFs have been taken at $ \mu=2 $ GeV as suggested in Refs.~\cite{Diehl:2004cx,Diehl:2007uc,Diehl:2013xca}.
Note also that the value of $ \alpha^{\prime} $ in the Eq.~(\ref{Eq:7}) and~(\ref{Eq:11}) has been set to $ \alpha^{\prime}=0.95 $ GeV$ ^{-2} $ which is in conformity with that which has been used in the study performed in Ref.~\cite{Diehl:2007uc} on the strange Dirac form factor $ F_1^{s} $.

According to the results obtained, one can conclude that if the ansatz Eq.~(\ref{Eq:11}) is used for calculating $ G_A $, the final results will not be remarkably sensitive to the choice of PPDFs set. To be more precise, according to the bottom panel of Fig.~(\ref{fig:fig1}), the difference between the results obtained for $ G_A $ using the \texttt{DSSV08} and \texttt{NNPDFpol1.1} PPDFs is almost less than 2\% in full range of $ Q^2 $, though the amounts of their uncertainties are somewhat different. However, Fig.~(\ref{fig:fig1}) clearly shows that the model fails to represent the data. As we shall show below, we can obtain an acceptable fit with a readjustment of parameters $\alpha'$ and $\mu$.

\subsection{Dependence on the scale \boldmath $ \mu $ of PPDFs}

Although the results presented in the previous subsection for the nucleon AFF $ G_A $ using the simple ansatz Eq.~(\ref{Eq:7}) follow roughly the experimental data, the question to be answered is to what extent the results will change if we take PPDFs at scales other than $ \mu=2 $ GeV. In Refs.~\cite{Diehl:2004cx}, the authors explained that the choice of scale should be a compromise between being large enough for PPDFs, $\Delta q_v(x)$, to be rather directly fixed by data and small enough to make contact with soft physics like conventional Regge phenomenology. However, since the recent analysis of PPDFs performed by the NNPDF Collaboration, namely \texttt{NNPDFpol1.1}~\cite{Nocera:2014gqa}, included a wide range of the available experimental data which covers the lower values of $ \mu $ down to $ \mu=1 $ GeV, it is also of interest to study the impact of taking PPDFs at a scale different from $ \mu=2 $ GeV on the theoretical predictions of $ G_A $ using the simple ansatz Eq.~(\ref{Eq:7}).

\begin{figure}[t!]
\centering
\includegraphics[width=0.53\textwidth]{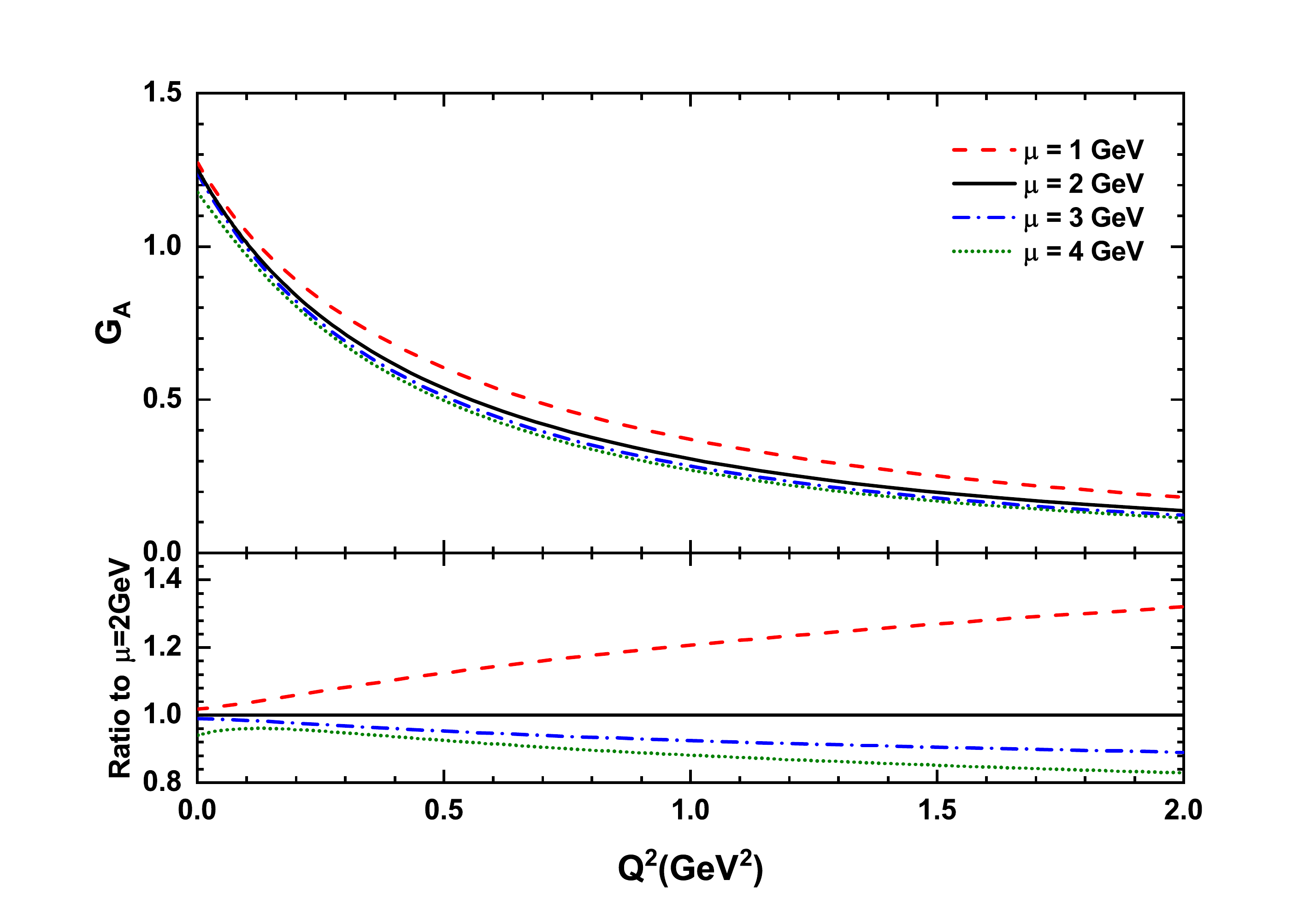}
\caption{The dependence of $ G_A $ on the value of scale $ \mu $ in which PPDFs are chosen. The calculations for the model have been performed using simple ansatz Eq.~(\ref{Eq:11})($\alpha'=0.95~\mathrm{GeV}^2$), NLO PPDFs of \texttt{NNPDFpol1.1}~\cite{Nocera:2014gqa} taken at $ \mu=1 $ (dashed), 2 (solid), 3 (dotted-dashed), and 4 (dotted) GeV. The ratios of the predictions to the corresponding result of $ \mu=2 $ GeV have been shown in the bottom panel.}
\label{fig:fig2}
\end{figure}

To evaluate the dependence of $ G_A $ on the value of scale $ \mu $ in which PPDFs are chosen, we repeat here the calculations performed in the previous subsection using the \texttt{NNPDFpol1.1} PPDFs but at different values of scale $ \mu $. The results obtained have been shown in Fig.~(\ref{fig:fig2}) where the dashed, solid, dotted-dashed and dotted curves correspond to \texttt{NNPDFpol1.1} PPDFs taken at $ \mu=1, 2, 3 $ and $ 4 $ GeV, respectively. In order to make a better comparison, in the bottom panel, we have also plotted the ratios of the predictions to the corresponding result obtained using \texttt{NNPDFpol1.1} PPDFs taken at $ \mu=2 $ GeV as a reference. As can be clearly seen, by decreasing the value of $ \mu $ in which PPDFs are chosen, $ G_A $ increases especially for larger values of $ Q^2 $, so that the difference between the results of $ \mu=1 $ and $ \mu=2 $ GeV reaches even to 30\% at $ Q^2=2 $ GeV$ ^2 $. On the other hand, as the value of $ \mu $ increases, $ G_A $ decreases but with a smaller rate than before, such that the difference between the results of $ \mu=2 $ and $ \mu=4 $ GeV reaches only to 20\% at $ Q^2=2 $ GeV$ ^2 $. Comparing the results of Figs.~(\ref{fig:fig1}) and~(\ref{fig:fig2}), one can conclude that taking PPDFs at a lower scale $ \mu $
can lead to a better description of the experimental data and lessen the relatively large discrepancy observed in Fig.~(\ref{fig:fig1}) between the predictions of the model and experimental data.

\subsection{Model uncertainties}

After studying the dependence of the nucleon axial form factor $ G_A $ on the PPDFs and also the value of scale $ \mu $ at which they are chosen, now we investigate the amount of uncertainties imposed on predictions of the model for $ G_A $ due to various sources and compare them with each other. According to sum rule given in Eq.~(\ref{Eq:8}), the model uncertainties in $ G_A $ can arise from the PPDFs uncertainties, the uncertainty of the scale $ \mu $ in which PPDFs are chosen, and the uncertainty of $ \alpha^{\prime} $ in the profile function $ f(x) $. We have studied the first two in Fig.~(\ref{fig:fig1}) and~\ref{fig:fig2}, respectively, and here investigate the uncertainties which arise from the $ \alpha^{\prime} $ variation. For this purpose, we repeat the calculations performed in Fig.~(\ref{fig:fig1}) using \texttt{NNPDFpol1.1} PPDFs~\cite{Nocera:2014gqa} taken at $ \mu=2 $ GeV, but this time vary $ \alpha^{\prime} $ in the range $ 0.85~\mathrm{GeV}^2 <\alpha^{\prime}< 1.15~\mathrm{GeV}^2 $.

\begin{figure}[t!]
\centering
\includegraphics[width=0.53\textwidth]{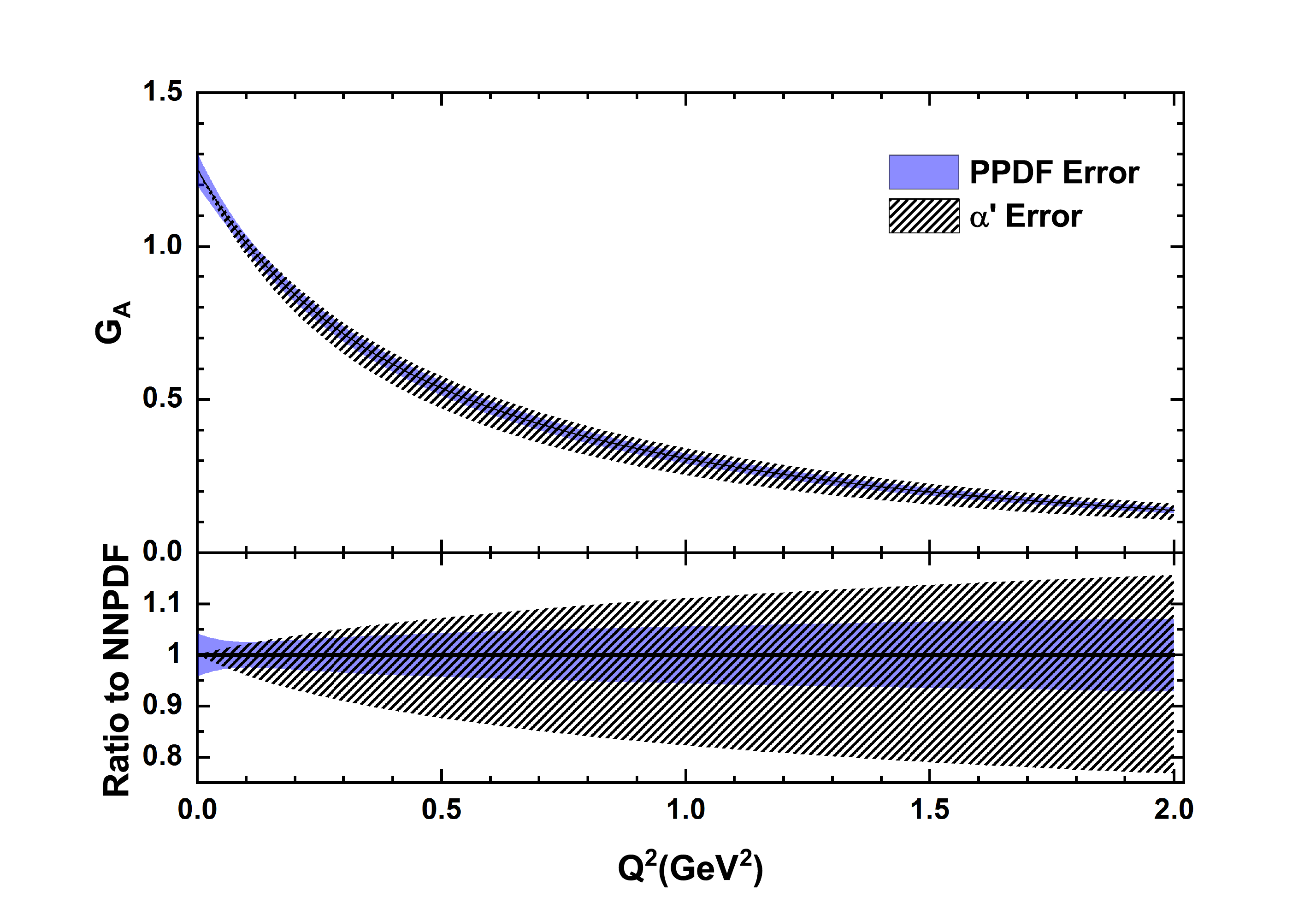}
\caption{A comparison between the model uncertainties in $ G_A $ due to the PPDFs uncertainties (filled band) and $ \alpha^{\prime} $ variations (hachured band) in the range $ 0.85~\mathrm{GeV}^2 <\alpha^{\prime}< 1.15~\mathrm{GeV}^2 $. The theoretical calculations have been performed using simple ansatz Eq.~(\ref{Eq:7}) with NLO PPDFs of \texttt{NNPDFpol1.1}~\cite{Nocera:2014gqa} taken at $ \mu=2 $ GeV. The bottom panel shows the relative uncertainties.}
\label{fig:fig3}
\end{figure}

Figure~(\ref{fig:fig3}) shows a comparison between the model uncertainties in $ G_A $ due to the PPDFs uncertainties (filled band) and $ \alpha^{\prime} $ variations (hachured band) in aforementioned range. The bottom panel shows the relative uncertainties obtained by dividing the upper and lower bands of each prediction by its central value. As can be seen, the uncertainty arising from the $ \alpha^{\prime} $ variations is remarkably dominant compared to the PPDFs uncertainty, except for very small values of $ Q^2 $ in which the PPDFs uncertainty becomes dominant. Note also that the uncertainty due to $ \alpha^{\prime} $ variations is asymmetric, while the PPDFs uncertainty is symmetric.
%

%
\section{\boldmath $ \chi^2 $ analysis of the experimental data}\label{sec:five} 

In the previous section, we found that the nucleon axial form factor $ G_A $ is not sensitive to the set of PPDFs chosen if an anzatz like Eq.~(\ref{Eq:11}) is used for connecting $ G_A $ to PPDFs via GPDs. However, according to the results obtained, any change in the scale $ \mu $ in which the PPDFs are taken and also the value of $ \alpha^{\prime} $ in profile function $ f(x) $ can lead to different results for $ G_A $. For this reason, in this section, we compute the best values for parameters of the model, i.e. $\mu$ and $\alpha'$, by performing $ \chi^2 $ analyses of the available experimental data. The optimization is done by
the CERN program \texttt{MINUIT}~\cite{James:1975dr}.

\subsection{Simple ansatz}

In oder to determine the best values of $ \mu $ and $ \alpha^{\prime} $ which are consistent with the experimental data of the nucleon axial form factor, various scenarios can be considered. As a first step, we perform a $ \chi^2 $ analysis of all data for the ratio $ G_A(Q^2)/G_A(0) $ from various experiments introduced in Sec.~\ref{sec:three}. For the theoretical calculations, we consider simple ansatz Eq.~(\ref{Eq:7}) with the \texttt{NNPDFpol1.1}~\cite{Nocera:2014gqa} as input PPDFs. Note that the theoretical calculation of quantity $ G_A(Q^2)/G_A(0) $ is not sensitive to the value of scale $ \mu $ in which PPDFs are chosen, since it is performed according to Eq.~(\ref{Eq:8}) and hence both the numerator and dominator include PPDFs similarly. We have examined various values of $ \mu $ and found that the results for the value of $ \alpha^{\prime} $ determined from the fit do not change up to four decimal places.

With 84 data points and 1 free parameter, the value of $ \chi^2 $ divided by the number of degrees of freedom is equal to $ \chi^2 /d.o.f=4.237 $. The value of  $ \alpha^{\prime} $ extracted from the fit is
\begin{equation}
\label{Eq:13}
\alpha^{\prime}= 2.754 \pm 0.0058 ~\textrm{GeV}^2,
\end{equation}
which is larger than the result obtained in Ref.~\cite{Diehl:2007uc} (about 1 GeV$ ^2 $). Using the reduced data set for $ G_A(Q^2)/G_A(0) $ which includes most precise point among the data points with the same $ Q^2 $ (40 data points), the value of $ \alpha^{\prime} $ is changed to $ \alpha^{\prime}= 2.476 \pm 0.0064 ~\textrm{GeV}^2 $. However, the value of $ \chi^2 /d.o.f $ increases to 5.129, since more than 40 data point with larger uncertainties have been removed from the analysis.  

Since, as mentioned before, the quantity $ G_A(Q^2)/G_A(0) $ cannot put any constraint on the value of scale $ \mu $ at which PPDFs are chosen, the above values obtained for $ \alpha^{\prime} $ are not very reliable. Actually, according to the results obtained in the previous section, we know that the change in $ \mu $ can change the result of $ G_A(Q^2) $ and subsequently the best value of $ \alpha^{\prime} $. Consequently, it is more reliable to extract the value of $\mu$ by performing a $ \chi^2 $ analysis of $ G_A(Q^2) $ data. For this purpose, we consider the reduced data for $ G_A(Q^2) $ that have been obtained from the original measurements of $ G_A(Q^2)/G_A(0) $, using $ G_A(0)= 1.2723\pm 0.0023 $~\cite{Tanabashi:2018oca}.

Since the quantity $ G_A(Q^2) $ is sensitive both to the value of $ \mu $ and $ \alpha^{\prime} $, we can determine their optimal values simultaneously. For this purpose, we can follow two procedures: 1) Performing several $ \chi^2 $ analysis by choosing different values for $ \mu $ as a fixed parameter and then minimizing $\chi^2$ with respect to $\alpha'$ and then plotting the $ \chi^2 $ as a function of $ \mu $ to find the point at which $ \chi^2 $ is an absolute minimum and its corresponding $ \alpha^{\prime} $. We call this procedure ``minimum tracing''. 2) Taking both $ \mu $ and $\alpha'$ as free parameters and minimizing $\chi^2$ with respect to both simultaneously. By following these two procedures, we can also find if there is a correlation between $ \mu $ and $ \alpha^{\prime} $.

\begin{figure}[t!]
\centering
\includegraphics[width=0.53\textwidth]{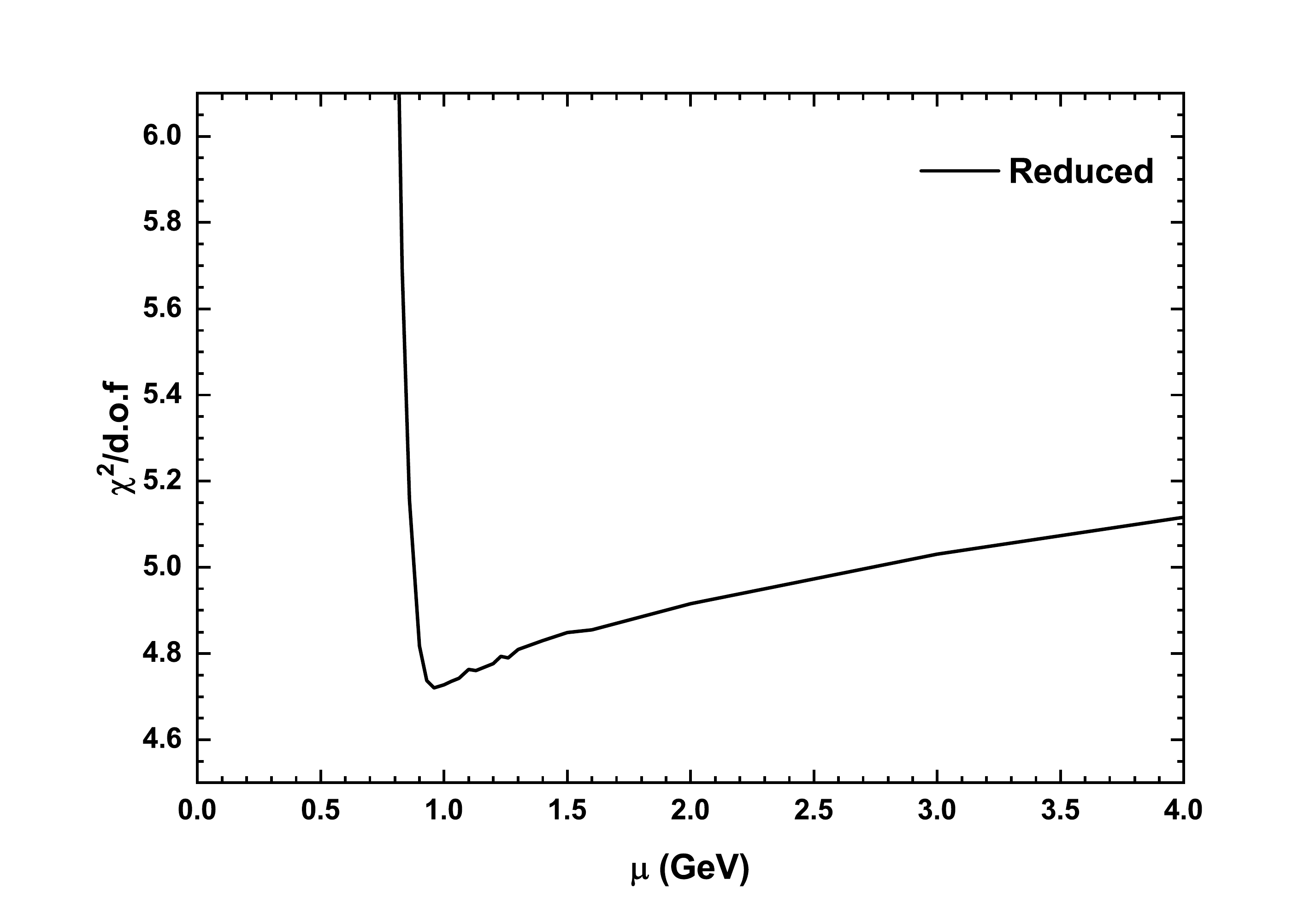}
\caption{The minimum tracing of $ \chi^2/d.o.f. $, shows the minimum of $ \chi^2/d.o.f. $  as a function of $ \mu $ in which PPDFs are chosen in calculation of the nucleon AFF Eq.~(\ref{Eq:8}), for the analysis of ``reduced" $ G_A(Q^2) $ data utilizing the simple ansatz for the profile function.}
\label{fig:fig4}
\end{figure}

Figure~(\ref{fig:fig4}) shows the results obtained for the minimum tracing of $ \chi^2/d.o.f.$ values of the reduced $ G_A(Q^2) $ data as a function of $ \mu $ in which PPDFs are chosen in calculation of the nucleon AFF Eq.~(\ref{Eq:8}). As can be seen, for the very small values of $ \mu $, the $ \chi^2 $ arises rapidly, while for the values greater than 1, it increases slowly. Note that the minimum occurs at about $ \mu= 0.96 $ GeV which is smaller than the value considered by the authors of Ref.~\cite{Diehl:2007uc} ($ \mu= 2 $ GeV). Moreover, the value of $ \alpha^{\prime} $ corresponding to this minimum is now as follows 
\begin{equation}
\label{Eq:14}
\alpha^{\prime}= 0.59 \pm 0.0014 ~\textrm{GeV}^2,
\end{equation}
which is also smaller than the result obtained in Ref.~\cite{Diehl:2007uc}.

As mentioned earlier, another method is to find the best values of $ \mu $ and $ \alpha^{\prime} $ from the $ \chi^2 $ analysis of reduced $ G_A(Q^2) $ data simultaneously. By performing such an analysis using \texttt{MINUIT}~\cite{James:1975dr}, the following results are obtained
\begin{align}
\label{Eq:15}
\alpha^{\prime}= 0.59 \pm 0.0022 ~\textrm{GeV}^2, \nonumber\\
\mu= 0.962 \pm 0.0098 ~\textrm{GeV},
\end{align}
which, as expected, are the same as those of the minimum tracing method. If we use all of the $ G_A(Q^2) $ data rather than its reduced set, these values are changed to
\begin{align}
\label{Eq:16}
\alpha^{\prime}= 0.65 \pm 0.0014 ~\textrm{GeV}^2, \nonumber\\
\mu= 0.987 \pm 0.11 ~\textrm{GeV}.
\end{align}
\begin{figure}[t!]
	\centering
	\includegraphics[width=0.45\textwidth]{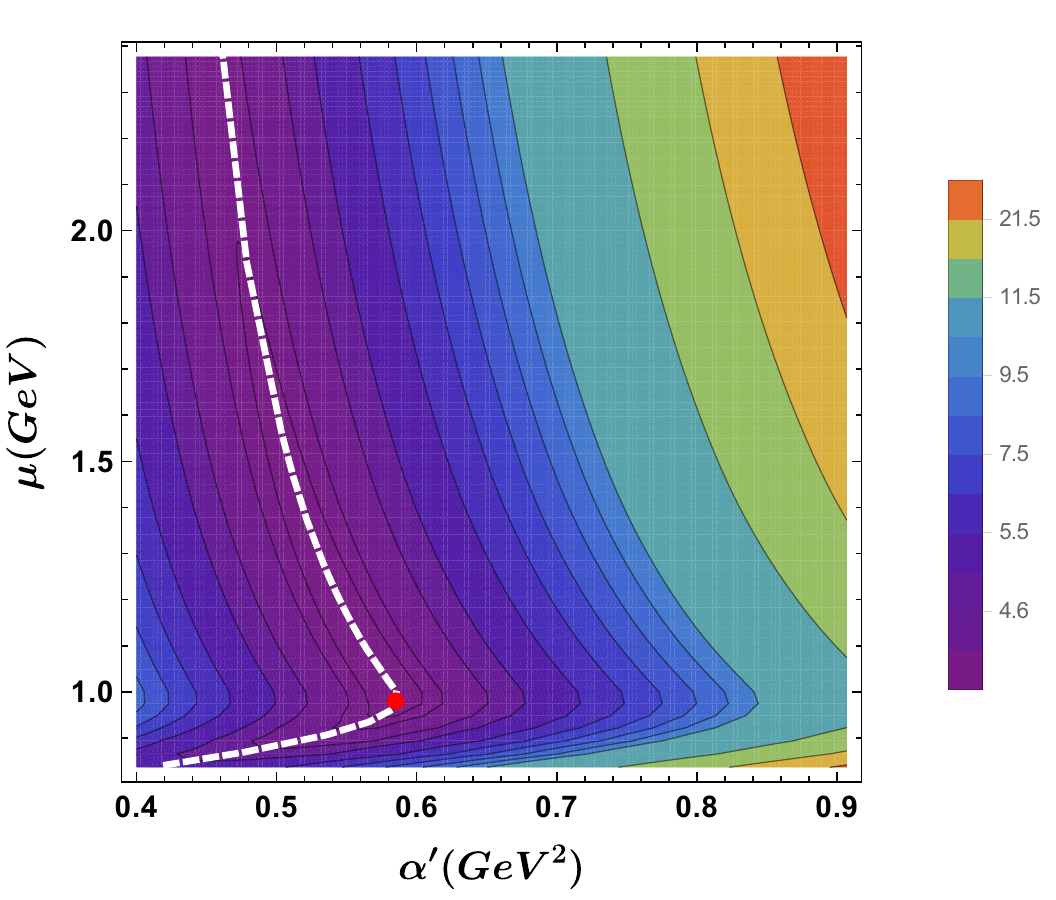}
	\caption{(Color online) Contour plot of $\chi^2/d.o.f.$ as a function of free parameters $ \mu $ and $ \alpha^{\prime} $ for the simple ansatz. The value of $\chi^2/d.o.f.$ is shown by colors (see bar legend on the right). The best value of $\chi^2/d.o.f.$ is shown as a red dot, the dashed line shows the path taken in Fig.~(\ref{fig:fig4}) for the minimum tracing procedure (see the text for more details).}
	\label{fig:fig5}
\end{figure}

The sensitivity of $ \chi^2 $ to the value of $ \mu $ and $ \alpha^{\prime} $ can be also studied in detail by plotting a contour plot which includes appropriate ranges of two parameters $ \mu $ and $ \alpha^{\prime} $. Figure~(\ref{fig:fig5}) shows the contour plot of $\chi^2/d.o.f.$ as a function of free parameters $ \mu $ and $ \alpha^{\prime} $. As can be seen from this figure, a steep rise in $\chi^2/d.o.f.$ occurs, if we increase or decrease $\alpha'$ more than 0.1 away from the best value. This fact means that the value of $\chi^2/d.o.f.$ is very sensitive to $\alpha'$, and that the reduced $ G_A(Q^2) $ data can put a good constraint on $\alpha'$. This is consistent with the small uncertainties for $\alpha'$ given in Eq.~(\ref{Eq:7})and~(\ref{Eq:16}). The situation is somewhat different for parameter $\mu$. Actually, a steep rise in the $\chi^2/d.o.f.$ occurs for $\mu > 2$ or $\mu < 0.9$, but for $0.9\lesssim\mu\lesssim 2$ one can see that $\chi^2/d.o.f.$ do not change as much. The minimum value of $\chi^2/d.o.f.$ has been shown as a red dot in the figure which occurs in $ \mu $ and $ \alpha^{\prime} $ values shown in Eq.~(\ref{Eq:15}). The dashed line shows the path in $\alpha'-\mu$ plane taken in Fig.~(\ref{fig:fig4}) for finding the minimum value of $\chi^2/d.o.f.$ using our minimum tracing procedure. In other words dashed line in Fig.~(\ref{fig:fig5}) shows the correspondance between the two procedures for finding the minimum of $\chi^2/d.o.f.$ explained earlier in this section; each point on this curve shows the pair $\alpha'$ and $\mu$ for which Fig.~(\ref{fig:fig4}) has a corresponding $\mu$ and $\chi^2/d.o.f.$ pair.

\begin{figure}[t!]
\centering
\includegraphics[width=0.53\textwidth]{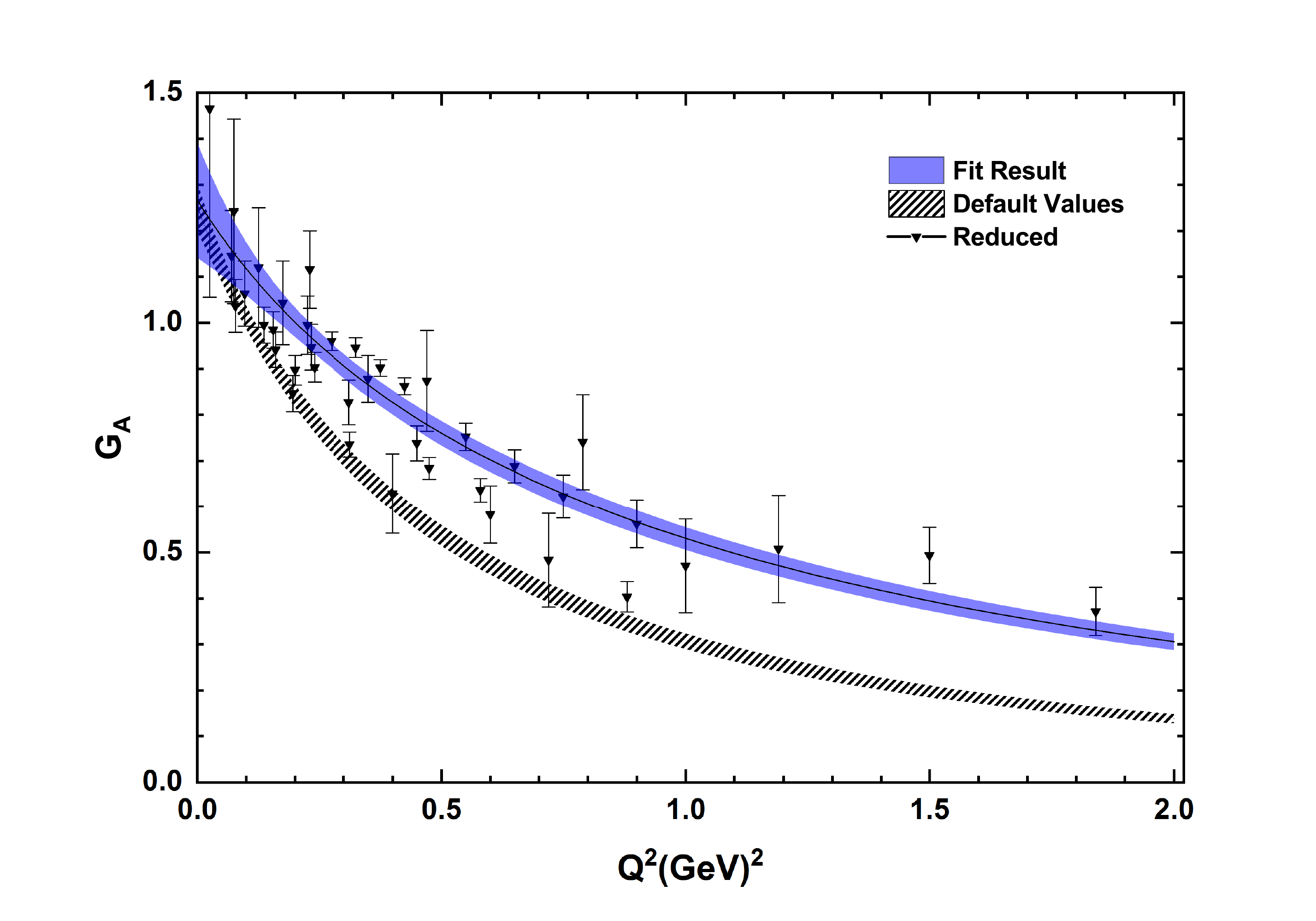}
\caption{A comparison between the reduced experimental data and theoretical predictions for $ G_A $ using the values obtained for $ \alpha^{\prime} $ and $ \mu $ (Eq.~(\ref{Eq:15})) from the the fit (filled band) and the results obtained using default values of ~\cite{Diehl:2007uc}, $ \alpha^{\prime}=0.95 $ GeV$ ^2 $ and $ \mu= 2 $ GeV (hachured band).}
\label{fig:fig6}
\end{figure}

Figure~(\ref{fig:fig6}) shows a comparison between the theoretical predictions for the nucleon axial form factor $ G_A $ using the values obtained for $ \alpha^{\prime} $ and $ \mu $ (Eq.~(\ref{Eq:15}))  from the analysis of reduced $ G_A(Q^2) $ data (filled band) and the results obtained using default values $ \alpha^{\prime}=0.95 $ GeV$ ^2 $ and $ \mu= 2 $ GeV (hachured band). The data points are those of the reduced set. As can be seen, the theoretical prediction is now in more consistent with the experimental data. 

\subsection{Complex ansatz}

Although Fig.~(\ref{fig:fig6}) clearly shows that using simple ansatz Eq.~(\ref{Eq:7}) can lead to an acceptable fit of the nucleon axial form factor $ G_A $ data, it is also of interest to investigate the effect of considering a more flexible profile function. In Ref.~\cite{Diehl:2004cx}, the authors indicated that low and high-$ x $ behavior of profile function $ f(x) $ and also the intermediate $ x $ region can be well characterized by the forms
\begin{equation}
\label{Eq:17}
f_q(x)=\alpha^{\prime}(1-x)^2\log\frac{1}{x}+B_q(1-x)^2 + A_qx(1-x),
\end{equation}
and
\begin{equation}
\label{Eq:18}
f_q(x)=\alpha^{\prime}(1-x)^3\log\frac{1}{x}+B_q(1-x)^3 + A_qx(1-x)^2.
\end{equation}
In this section, we examine these profile functions to see whether any improvement in the theoretical predictions and the fit can be achieved. Note that for the calculation of $ G_A $ according to Eq.~(\ref{Eq:8}), one in principle  needs to consider profile function Eq.~(\ref{Eq:17}) (or Eq.~(\ref{Eq:18})) for each flavor $ u_v $, $ d_v $, $ \bar u $ and $ \bar d$. In this way, there are 8 more free parameters that should be determined from the analysis of $ G_A $ data. The best procedure for selecting the most appropriate parameters and then finding the optimal parametrization form is performing a parametrization scan as described in Ref.~\cite{Aaron:2009aa} for the case of PDFs determination through a QCD analysis of HERA DIS data.

First, we consider the profile function Eq.~(\ref{Eq:17}) and again use the reduced set of $ G_A $ data. The value of $ \mu $ in which PPDFs are chosen is set to the value obtained in Eq.~(\ref{Eq:15}) using simple ansatz Eq.~(\ref{Eq:7}). By performing a parametrization scan, it is found that none of the free parameters can lead to a decrease in the value of $ \chi^2 $ more than one unit as compared to the corresponding value for the analysis using the simple ansatz (see previous subsection). Consequently, adding some free parameters in the form of Eq.~(\ref{Eq:17}), even for the valence quark profile functions, will not have any effect on the fit quality. However, if we use the profile function Eq.~(\ref{Eq:18}) instead, some improvements can be achieved in the fit quality. We find that the only parameters that can lead to a significant decrease in the value of $ \chi^2 $ are the valence quarks parameters. Moreover, by considering $ A_{u_v} $, $ B_{u_v} $, $ A_{d_v} $ and $ B_{d_v} $ as free parameters and setting the other parameters for $ \bar u $ and $ \bar d $ equal to zero, the value of $ \chi^2 $ decreases from $ 184.1 $ (which is $ \chi^2 $ for the analysis of reduced $ G_A $ data using simple ansatz Eq.~(\ref{Eq:7})) to $ 173.6 $. Next we investigate the possibility of assuming $ u_v $ and $ d_v $ parameters to be equal, to reduce the number of free parameter as much as possible without damaging the quality of the fit. For this purpose, we consider $ A_{u_v}=A_{d_v}=A_v $ and $ B_{u_v}=B_{d_v}=B_v $, so that only two extra parameters contribute in the fit. The value of $ \mu $ is again set to the value obtained in Eq.~(\ref{Eq:15}). As a result, we find that the value of $ \chi^2 $ changes less than two units, namely from $ 173.6 $ to $ 175.0 $. It means that it is acceptable to take the $ u_v $ and $ d_v $ parameters to be equal and reduce the number of free parameters. Then, the optimal values obtained for the parameters of the fit are as follows
\begin{equation}
\label{Eq:19}
\begin{split}
& \alpha^{\prime}= 1.029 \pm 0.22 ~\textrm{GeV}^2, \\
& A_v= 12.74 \pm 2.20,~~~~~~ B_v= -3.5 \pm 0.64.
\end{split}
\end{equation}
As can be seen, the value of $ \alpha^{\prime} $ now has increased to about $1.0$ which is consistent with the result obtained in Ref.~\cite{Diehl:2007uc}.

\begin{figure}[t!]
\centering
\includegraphics[width=0.53\textwidth]{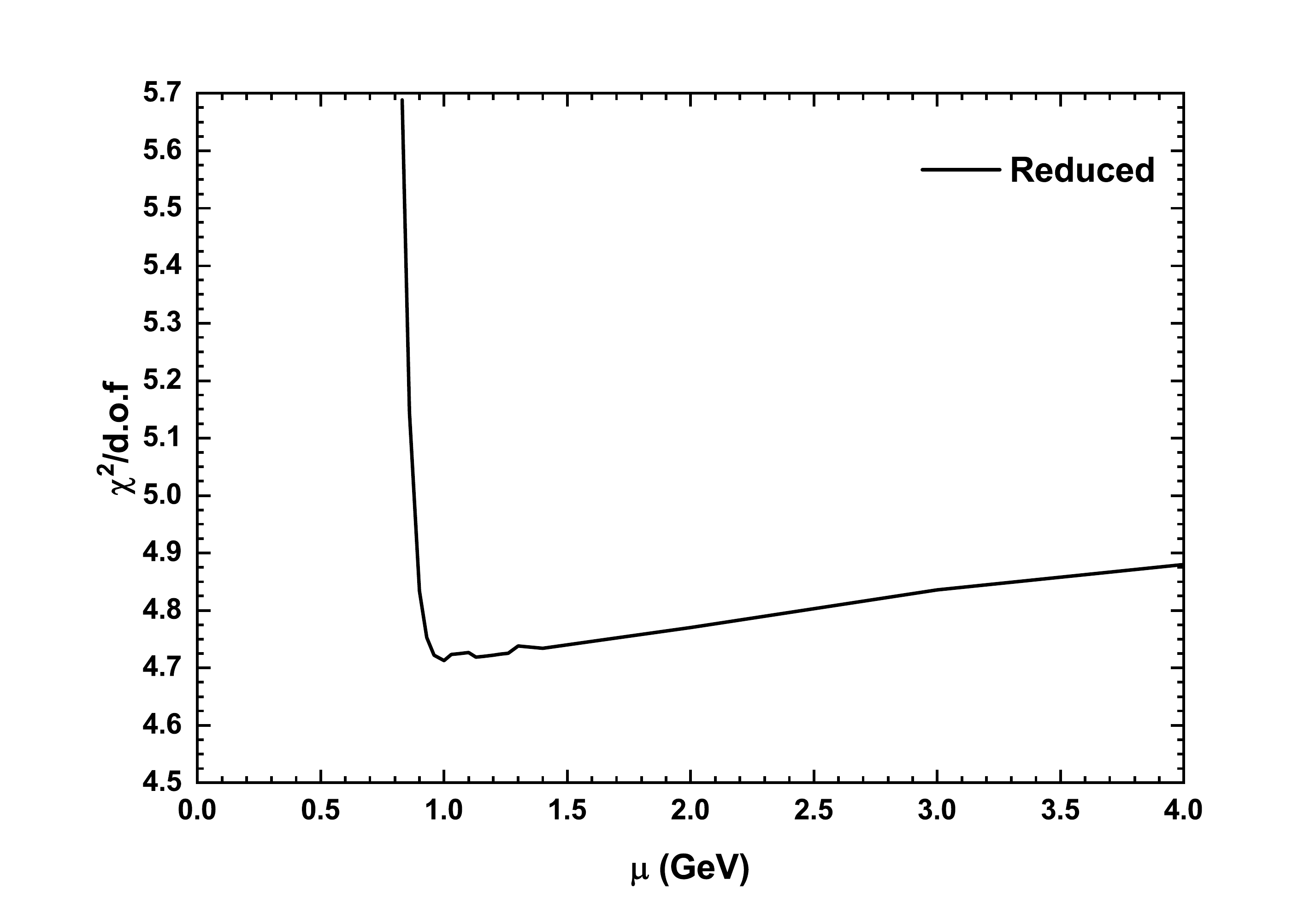}
\caption{Same as Fig.~(4), but using Eq.~(\ref{Eq:18}) for the profile function.}
\label{fig:fig7}
\end{figure}

For the analyses performed so far in this subsection, we have set the value of $ \mu $ in which PPDFs are chosen equal to $ \mu= 0.962 $ GeV according to Eq.~(\ref{Eq:15}). However, we should find the best value of $ \mu $ by performing a minimum tracing or considering it as a free parameter of the fit just like the previous section. Figure~(\ref{fig:fig7}) shows the results obtained by minimum tracing of $ \chi^2/d.o.f.$ values of the reduced $ G_A(Q^2) $ data as a function of $ \mu $ using the profile function Eq.~(\ref{Eq:18}) with same $ A_v $ and $ B_v $ parameters for valence quarks and setting the corresponding sea quark parameters equal to zero. According to this figure the minimum occurs at about $ \mu= 1.0 $ GeV which is somewhat larger than before, but still smaller than $ \mu= 2.0 $ GeV which has been considered in Ref.~\cite{Diehl:2007uc}. In this situation, the optimal values obtained for the parameters of the fit are as follows
\begin{equation}
\label{Eq:20}
\begin{split}
& \alpha^{\prime}= 1.054 \pm 0.22 ~\textrm{GeV}^2, \\
& A_v= 13.28 \pm 2.00,~~~~~~ B_v= -3.64 \pm 0.64.
\end{split}
\end{equation}
Next we examine the effect of considering $ \mu $ as a free parameter whose value is to be determined by simultaneous optimization along with the other three parameters. We find that the results do not change significantly, just like before. The results are,
\begin{equation}
\label{Eq:21}
\begin{split}
& \alpha^{\prime}= 1.054 \pm 0.22 ~\textrm{GeV}^2,~~\mu=0.997 \pm 0.363 ~\textrm{GeV}\\
& A_v= 13.28 \pm 2.00,~~~~~~~~~ B_v= -3.64 \pm 0.64.
\end{split}
\end{equation}
All things considered, we can conclude that using reduced set of $ G_A $ data to determine the best value of $ \mu $, in which PPDFs are chosen, leads to a smaller amount for it (about $ \mu= 1.0 $ GeV) as compared to the value assumed in Ref.~\cite{Diehl:2007uc}, whether a simple ansatz is used or a more flexible ansatz like Eq.~(\ref{Eq:18}). However, for the case of $ \alpha^{\prime} $, the situation is somewhat different. Actually, using simple ansatz Eq.~(\ref{Eq:7}) leads to $ \alpha^{\prime}=0.59 $ which is smaller than the one obtained by the study of strange Dirac form factor $ F_1^{s} $~\cite{Diehl:2007uc}, while using a complex ansatz like Eq.~(\ref{Eq:18}) leads to a value about $ \alpha^{\prime}= 1.054 ~\textrm{GeV}^2 $ which is in consistent with the result of Ref~\cite{Diehl:2007uc}. 

\begin{figure}[t!]
\includegraphics[width=0.5\textwidth]{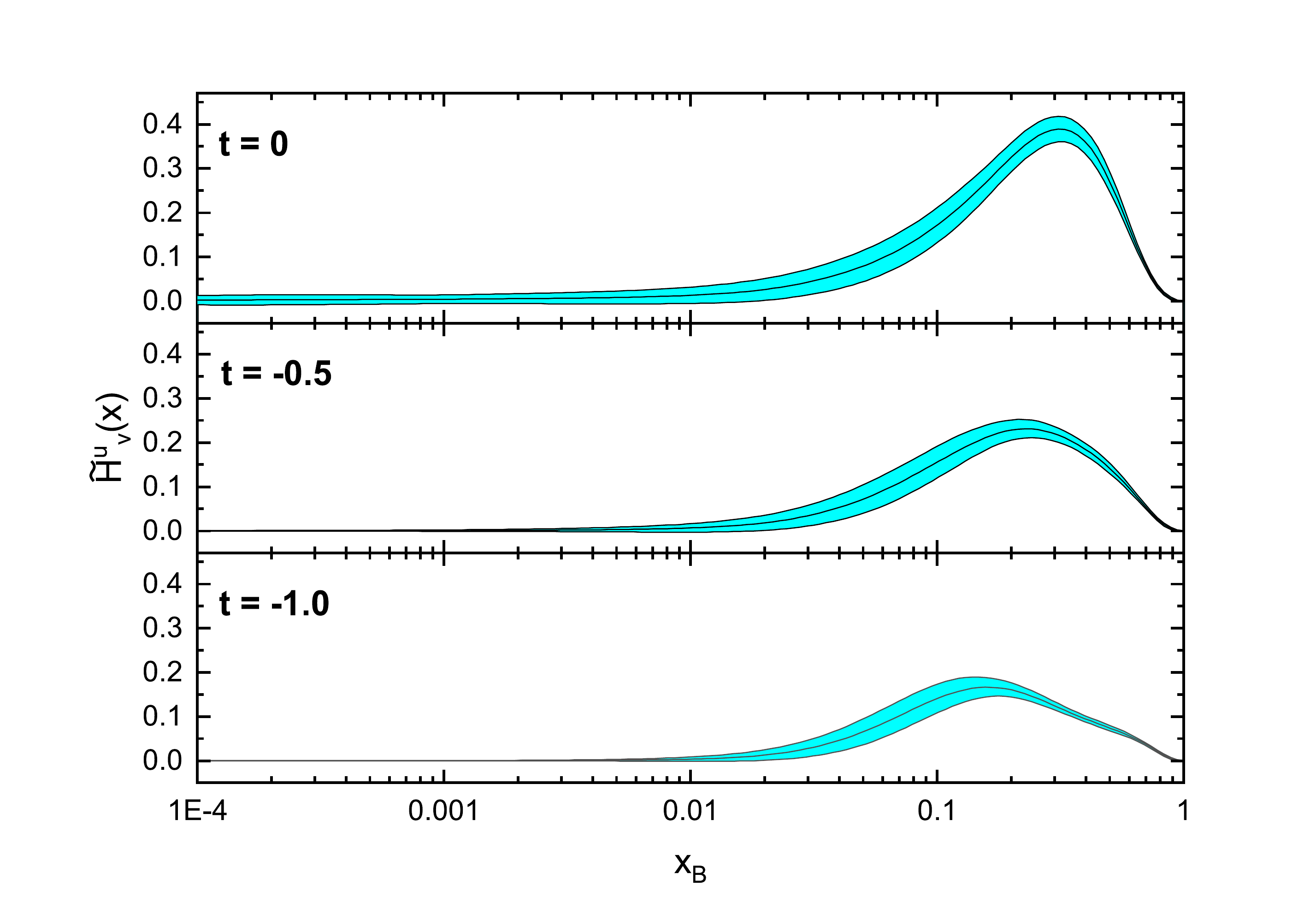}
\includegraphics[width=0.5\textwidth]{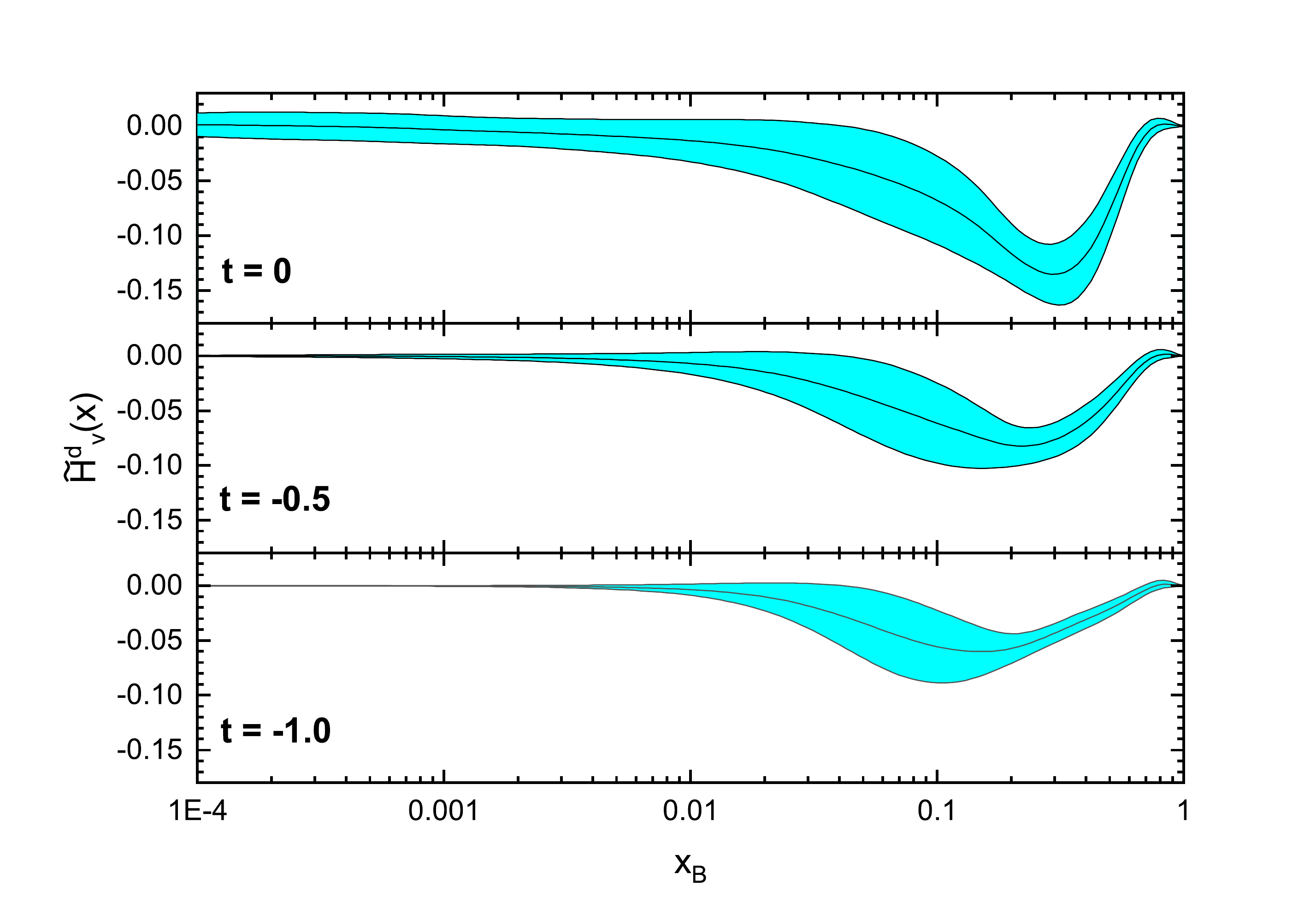}
\caption{The polarized GPDs $ \widetilde{H}_{u_v} $ (top) and $ \widetilde{H}_{d_v} $ (bottom) as a function of $ x $ at $ \mu=1 $ GeV for three different values of $ t=-Q^2= 0, -0.5 $. The theoretical calculations have been performed using ansatz Eq.~(\ref{Eq:11}) with profile function Eq.~(\ref{Eq:18}) and values Eq.~(\ref{Eq:21}).}
\label{fig:fig8}
\end{figure}
\begin{figure}[t!]
\centering
\includegraphics[width=0.5\textwidth]{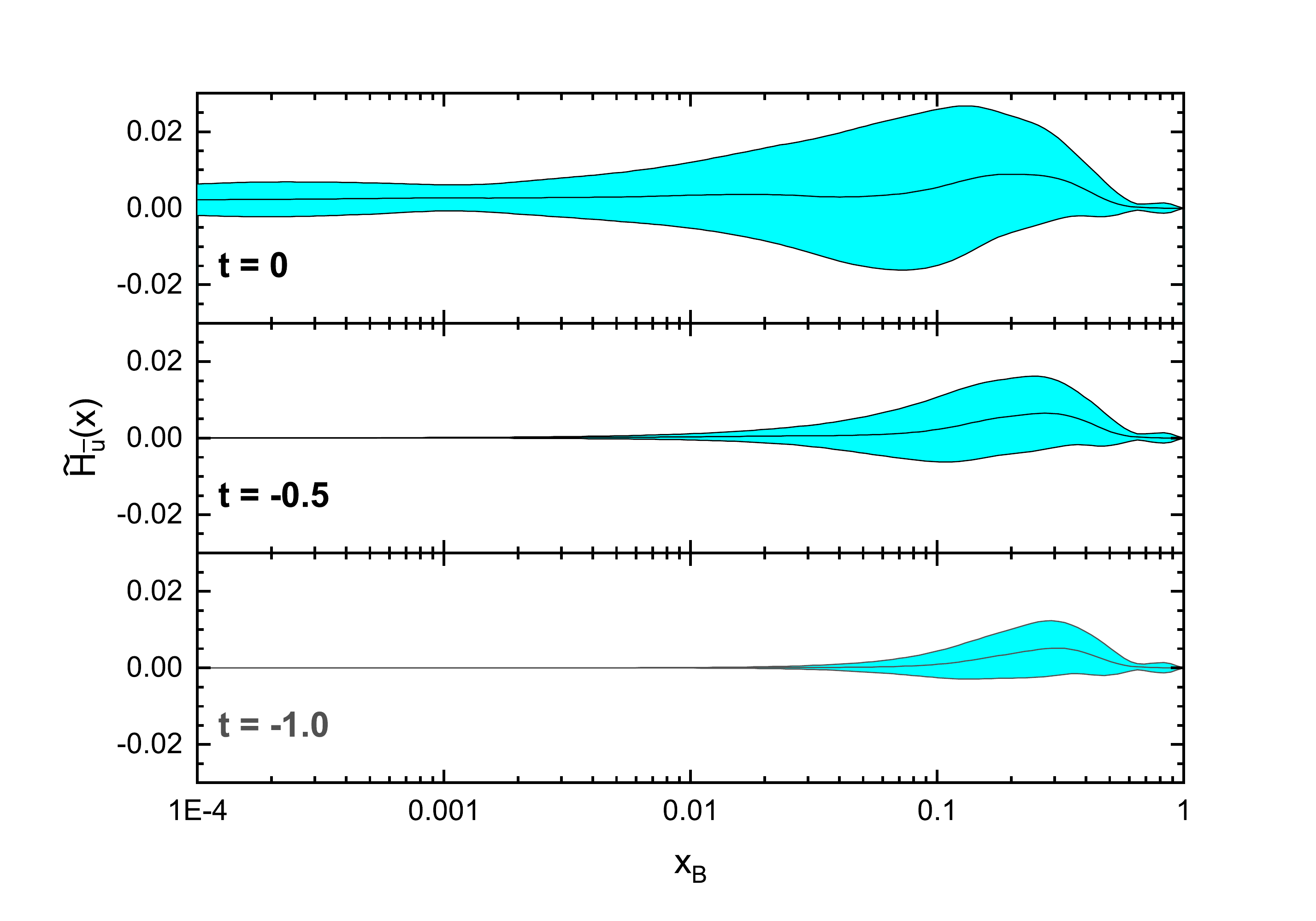}
\includegraphics[width=0.5\textwidth]{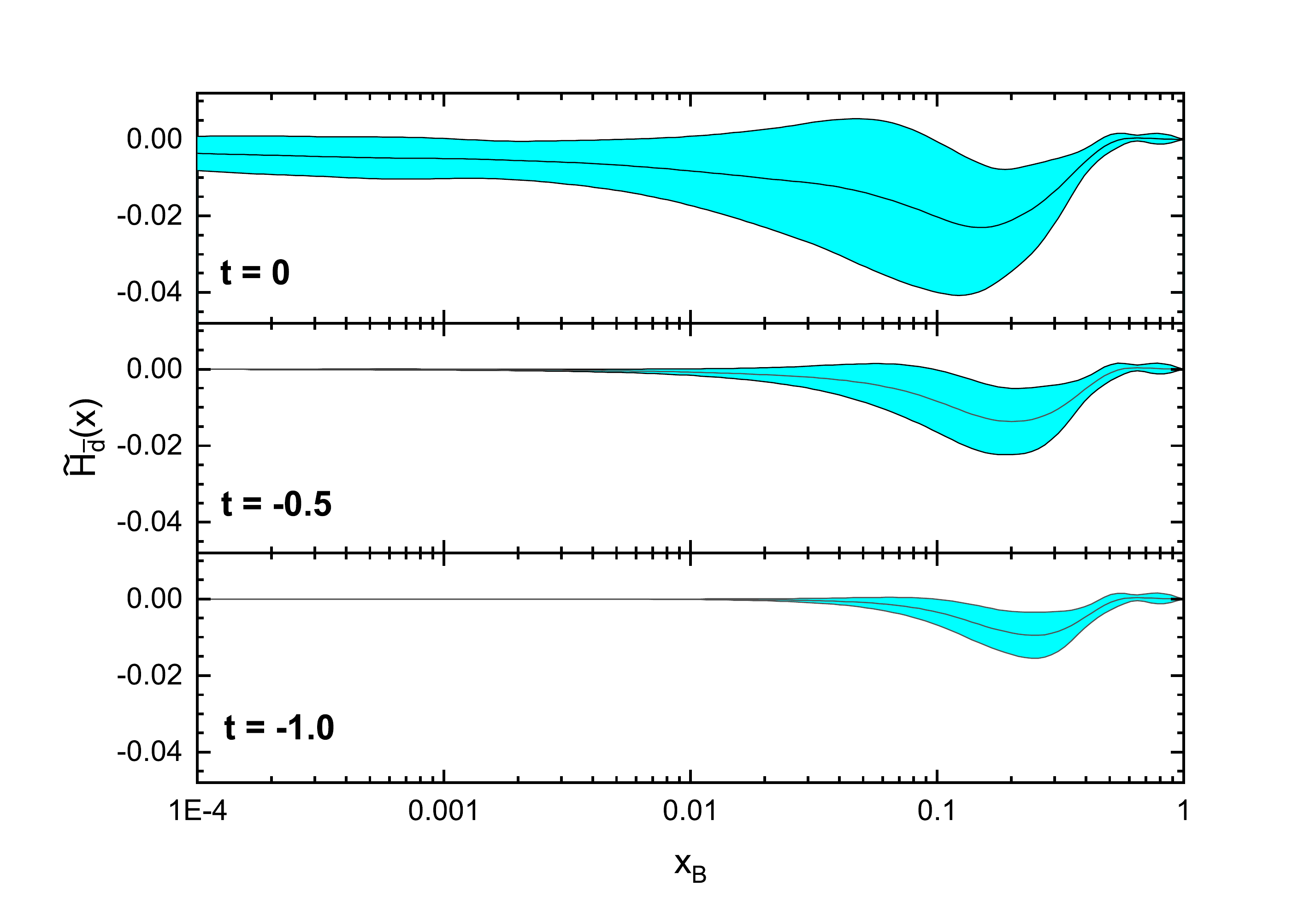}
\caption{Same as Fig.~(\ref{fig:fig8}), but for $ \widetilde{H}_{\bar u} $ and $ \widetilde{H}_{\bar d} $.}
\label{fig:fig9}
\end{figure}

It is also of interest now to plot polarized GPDs according ansatz Eq.~(\ref{Eq:11}) and using profile function Eq.~(\ref{Eq:18}) and values shown in Eq.~(\ref{Eq:21}) obtained from final analysis. Figure~(\ref{fig:fig8}) shows polarized GPDs $ \widetilde{H}_{u_v} $ (top) and $ \widetilde{H}_{d_v} $ (bottom), as a function of $ x $ for three different values of $ t=-Q^2= 0, -0.5 $ and 1 GeV$ ^2 $. Note that for $ t=0 $, we obtain the original polarized PDFs from \texttt{NNPDFpol1.1}~\cite{Nocera:2014gqa}. As can be seen, as the absolute value of $ t $ increases, the distributions for the valence quarks decrease in magnitude and shift somewhat to smaller values of $ x $, as expected. Note that, the uncertainty of $ \widetilde{H}_{u_v} $ is less than $ \widetilde{H}_{d_v} $, since $ \Delta u_v$ PPDF of \texttt{NNPDFpol1.1} has less uncertainty. The corresponding plots for $ \widetilde{H}_{\bar u} $ and $ \widetilde{H}_{\bar d} $ are shown in Fig~(\ref{fig:fig9}). Note that in this case, the parameters $ A $ and $ B $ in profile function Eq.~(\ref{Eq:18}) are equal to zero. This figure shows that as the absolute value of $t$ increases, the distributions for the sea quarks slightly decrease in magnitude and shift to larger $x$. However their uncertainty bands are greater than those of the valence quarks. 

\begin{figure}[t!]
\centering
\includegraphics[width=0.5\textwidth]{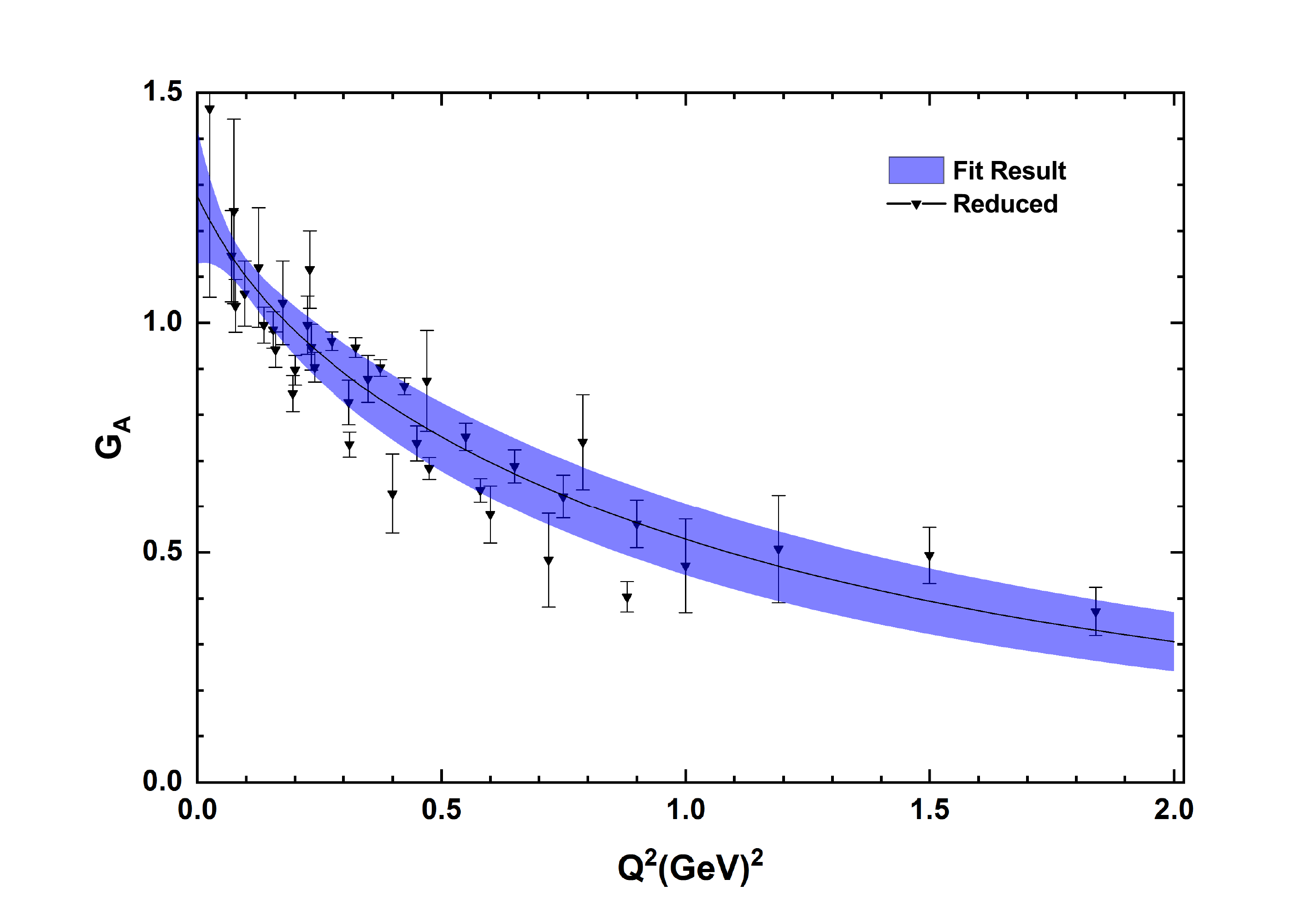}
\caption{A comparison between the reduced experimental data and theoretical predictions for $ G_A $ obtained using profile function Eq.~(\ref{Eq:18}) with values of parameters given Eq.~(\ref{Eq:21}) (filled band). \texttt{NNPDFpol1.1} PPDFs have been chosen at $ \mu=0.997 $ GeV.}
\label{fig:fig10}
\end{figure}

We can now compare the result of our final analysis to the experimental data. Figure~(\ref{fig:fig10}) shows a comparison between the theoretical predictions for the nucleon AFF $ G_A $ obtained using profile function Eq.~(\ref{Eq:18}) with values Eq.~(\ref{Eq:20}) (filled band), and reduced $ G_A(Q^2) $ data. Note that \texttt{NNPDFpol1.1} PPDFs have been chosen at $ \mu=0.997 $ GeV as shown in Eq.~(\ref{Eq:21}). As can be seen, the theoretical prediction is in a good agreement with the experimental data.

%
 
%
\section{Summary and Conclusions}\label{sec:six}

An accurate knowledge of generalized parton distributions is necessary for describing hard exclusive electroproduction processes. GPDs are non-perturbative objects which can be determined using analysis of experimental data from exclusive processes such as DVCS and meson production. One of the important properties of GPDs is their mutual relations with PDFs and form factors. To be more precise, (polarized) GPDs reduce to (polarized) PDFs in the limit of zero momentum transfer. On the other hand, the integration of GPDs over Bjorken $ x $ yields Dirac and Pauli form factors. This procedure for the case of polarized GPDs yields AFF or intrinsic quark contribution to the nucleon spin. In the present work, considering Diehl's model~\cite{Diehl:2004cx,Diehl:2007uc,Diehl:2013xca} to relate GPDs and PDFs, we have calculated polarized GPDs ($\widetilde{H}_q$) using predetermined polarized PDFs and have studied in details the axial form factor of nucleon $ G_A $. As a result, we have shown that our model to calculate $ G_A $ is not sensitive to the choice of PPDFs set, such that the difference between the results obtained using the \texttt{DSSV08} and \texttt{NNPDFpol1.1} PPDFs is almost less than 2\% in full range of $ Q^2 $. By studying the dependence of $ G_A $ on the value of scale $ \mu $ at which PPDFs are chosen, we can also found that as $ \mu $ decreases, $ G_A $ increases especially for larger values of $ Q^2 $, so that the difference between the results of $ \mu=1 $ and $ \mu=2 $ GeV reaches 30\% at $ Q^2=2 $ GeV$ ^2 $. Overall, we have concluded that taking PPDFs at a lower scale $ \mu $
can lead to a better description of the experimental data. Moreover, we have investigated the model uncertainties imposed on $ G_A $ due to the PPDFs uncertainties and also variation of $ \alpha^{\prime} $ in profile functions $ f(x) $. We have indicated that the uncertainty arising from the $ \alpha^{\prime} $ variations is dominant as compared to the PPDFs uncertainty, except for very small values of $ Q^2 $ in which the PPDFs uncertainty becomes dominant. Moreover, by considering different scenarios, we have determined the optimal values of parameters of the model using standard $\chi^2$ analysis of the available experimental data related to nucleon axial form factor. We used both a simple and complex profile function to find the best conditions for obtaining better consistency between the theoretical predictions and experimental data. We have shown that using $ G_A $ data to determine the best value of $ \mu $, in which PPDFs are chosen, leads to a smaller amount for it (about $ \mu= 1.0 $ GeV) compared with the value assumed in Ref.~\cite{Diehl:2007uc} whether one uses a simple ansatz or a more flexible ansatz. In addition, using a simple ansatz leads to a smaller value for $ \alpha^{\prime} $ rather than one obtained by the study of strange Dirac form factor $ F_1^{s} $~\cite{Diehl:2007uc}, while using a complex ansatz leads to $ \alpha^{\prime}= 1.054 ~\textrm{GeV}^2 $ which is in consistent with the result of Ref~\cite{Diehl:2007uc}.
More precise measurements of neutrino cross section on hydrogen and deuterium are needed to unravel the axial structure of the nucleon.

%
\section*{ACKNOWLEDGEMENT}
We thank H. Abedi and M. J. Kazemi for reading the manuscript and for stimulating discussions. We also wish to express our gratitude toward U. G. Meissner and A. Butkevich for providing us with the data. HH and SSG would like to thank the research council of the Shahid Beheshti University for financial support. MG thanks the School of Particles and Accelerators, Institute for Research in Fundamental Sciences (IPM) for financial support provided for this research.

%

\end{document}